\documentclass[journal,10pt,twocolumn]{IEEEtran}
\usepackage{epsfig}
\usepackage{graphicx}
\usepackage{subfigure}

\usepackage{amssymb}
\usepackage{makeidx}
\usepackage{amsmath}
\usepackage{graphicx}
\usepackage{epsf}
\usepackage{epsfig}
\usepackage{psfig}
\usepackage{ccaption}
\usepackage{array}
\usepackage{tabularx}
\usepackage{multirow}
\usepackage{epsfig}
\usepackage{cite}
\usepackage{procedure,algorithm,algorithmic}

\begin{document}

\title{Combating False Reports for Secure Networked Control in Smart Grid via Trustiness Evaluation}
\author{Husheng Li, Lifeng Lai and Seddik M. Djouadi
\thanks{H. Li and S. Djouadi are with the Department of Electrical Engineering and Computer Science, the University of Tennessee, Knoxville, TN, 37996 (email:
husheng@eecs.utk.edu, djouadi@utk.edu). L. Lai is with the Department of Systems Engineering, University of Arkansas, Little Rock, AR, (email: lxlai@ualr.edu). This work was supported by the National
Science Foundation under grants CCF-0830451 and ECCS-0901425.}}

\maketitle

\begin{abstract}
Smart grid, equipped with modern communication infrastructures, is subject to possible cyber attacks. Particularly, false report attacks which replace the sensor reports with fraud ones may cause the instability of the whole power grid or even result in a large area blackout. In this paper, a trustiness system is introduced to the controller, who computes the trustiness of different sensors by comparing its prediction, obtained from Kalman filtering, on the system state with the reports from sensor. The trustiness mechanism is discussed and analyzed for the Linear Quadratic Regulation (LQR) controller. Numerical simulations show that the trustiness system can effectively combat the cyber attacks to smart grid.
\end{abstract}

\section{Introduction}
\begin{figure}
  \centering
  \includegraphics[scale=0.6]{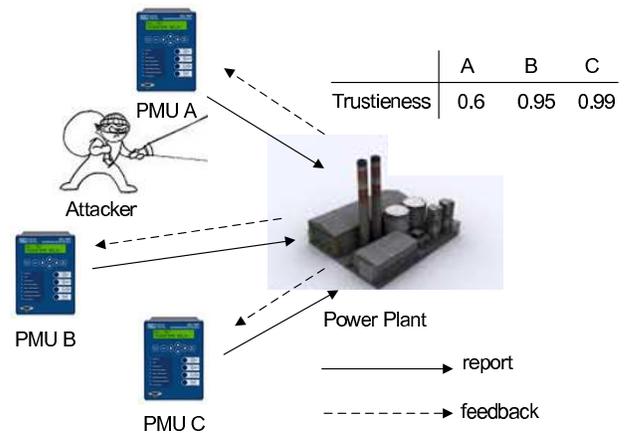}
  \caption{An illustration of the secure control system in smart grid.}\label{fig:illu}
\end{figure}

In recent years, smart grid has attracted significant interest in both communities of communications and power systems \cite{ISO2009}. In a smart grid, modern communication technologies are used to convey information like system parameters (voltage, frequency, harmonics, etc) and power consumption information in order to improve the robustness, agility and efficiency of power grid. For example, as illustrated in Fig. \ref{fig:illu}, phasor measurement units (PMUs) send report to the power plant which takes actions to stabilize the power grid.

However, the communication infrastructure also brings vulnerability to the smart grid. An attacker can attack the communication links using various approaches such as jamming in the physical layer and Byzantine attack in the upper layers. The attack could result in delay or drop of report packets. The attacker may revise the report such that the received report is wrong, thus possibly incurring system instability and even large area blackout which brings the loss of millions of dollars. Therefore, a secure design of smart grid is in a pressing need.

In this paper, we study the trustiness framework based secure control in smart grid. We assume that each report could be substituted by a false report from an attacker\footnote{For the packet delay or loss, many techniques have been developed in the area of networked control, e.g., the LQR control subject to packet losses \cite{Matveev2009}.}. Instead of studying the security protocols in the communication networks, which has been intensively studied for data networks like Internet, we focus on the controller side, i.e., the power plant, which is aware of possible attacks. The controller can predict the future system state and then evaluate the trustiness of reports from different PMUs. Based on these trustiness, the controller takes a corresponding control strategy, e.g., dropping untrusted packets. The trustiness will also be fed back to the PMUs such that the PMUs can adjust their security setups (e.g., the keys or the cryptography protocols). The system is illustrated in Fig. \ref{fig:illu}. For example, the reports from PMU A is attacked by the attacker. Then, the trustiness of the reports from PMU A is reduced, which may make the controller drop the corresponding reports. Meanwhile, PMU A may change its key or use a more secure protocol when it finds out that its trustiness has been significantly decreased.

Note that the reliability issues for control systems have been considered in supervisory control and data acquisition (SCADA) standard \cite{SCADA2007}. However, SCADA is mostly focused on the reliability subject to random failures, instead of malicious attacks. There are some studies on the control systems subject to malicious attacks \cite{Amin2009}\cite{Cardenas2008_1}\cite{Cardenas2008_2}. \cite{Cardenas2008_1} and \cite{Cardenas2008_2} introduce general problems and approaches for secure control without exploring the details. \cite{Amin2009} is focused on the control system subject to denial-of-service (DoS) attacks, thus mainly addressing the packet delays or losses. There have not been any studies on combating the proofing attacks, particularly, applying the trustiness framework in the secure control.

Note that the trustiness system has been widely used in information systems, e.g., \cite{Josang2007}\cite{Stumpt2007}. However, they are not designed for control systems. The unique system dynamics of the control system can be exploited to build the corresponding effective trustiness. In this paper, we apply Kalman filtering \cite{Poor1994} to predict the system state using different combinations of PMU reports, thus realizing a cross-check of the report trustiness.

The remainder of this paper is organized as follows. The system model of the controller and communication infrastructure is introduced in Section \ref{sec:system}. A mechanism of evaluating the trustiness of different PMUs is proposed in Section \ref{sec:evaluation}. The numerical simulation results and conclusions are provided in Sections \ref{sec:numerical} and \ref{sec:conclusion}.

\section{System Model}\label{sec:system}
In this section, we first introduce the linear system model for power grid. Then, we explain the LQR criterion of the control.

\subsection{Linear System}
We model the dynamics of power grid as a discrete-time linear system\footnote{Although power grid is usually nonlinear, it can be approximated by a linear system in the small perturbation case.}, whose dynamics are given by
\begin{eqnarray}
\mathbf{x}(t+1)&=&\mathbf{A}\mathbf{x}(t)+\mathbf{B}\mathbf{u}(t)+\mathbf{w}(t),\nonumber\\
\mathbf{y}(t)&=&\mathbf{C}\mathbf{x}(t)+\mathbf{n}(t),
\end{eqnarray}
where $\mathbf{x}$ is an $N$-vector and represents the system state of the power grid, $\mathbf{y}$ is an $M$-vector representing the observations and $\mathbf{u}$ is the action taken by the controller. The matrices $\mathbf{A}$, $\mathbf{B}$ and $\mathbf{C}$ are specified by the system. Both $\mathbf{w}(t)$ and $\mathbf{n}(t)$ are Gaussian noise. For simplicity, we suppose that each dimension of the observation vector $\mathbf{y}$ is sensed by a PMU. It is easy to extend to the general case in which the sensor reports have overlaps. Each PMU sends its observations to the controller via a communication channel since they are not located at the same place as the controller.

We use the following assumptions throughout the paper:
\begin{itemize}
\item Each report can be successfully received by the controller if there is no attack. This is reasonable for communication channels with good qualities. For the practical case of occasional packet drop, the control strategy can be obtained by considering the corresponding element in the observation matrix $\mathbf{C}$ as zero.

\item Each report could be replaced with a false report by an intervening attacker. The attacker could intercept the report and insert its own one. However, we assume that not all reports are replaced.

\item For simplicity, we assume that there is at most one attacker. The principle of the trustiness system can be extended to the case of multiple attackers at the cost of more computational cost.
\end{itemize}

\subsection{LQR Control}
When there is no attacker, we assume that the controller adopts the LQR control \cite{Kwakernaak1972} with an infinite time horizon with the cost function given by
\begin{eqnarray}\label{eq:LQG}
J=E\left[\sum_{t=1}^{\infty}\beta^t\left(\mathbf{x}^T(t)\mathbf{Q}\mathbf{x}(t)+\mathbf{u}^T(t)\mathbf{P}\mathbf{u}(t)\right)\right],
\end{eqnarray}
where $\mathbf{Q}$ and $\mathbf{P}$ are positive definite matrices. The physical meaning of the objective function is given below:
\begin{itemize}
\item The term $\mathbf{x}^T(t)\mathbf{Q}\mathbf{x}(t)$ is the norm (with respect to the positive definite matrix $\mathbf{Q}$) of the system state vector.

\item The term $\mathbf{u}^T(t)\mathbf{P}\mathbf{u}(t)$ is the norm of the action vector with respect to the matrix $\mathbf{P}$, which represents the cost due to the action itself.
\end{itemize}

Based on the cost function in (\ref{eq:LQG}), the LQR action $\mathbf{u}(t)$ is given by
\begin{eqnarray}
\mathbf{u}(t)=-\mathbf{L}\hat{x}(t),
\end{eqnarray}
where $\hat{x}(t)$ is the estimation of the system state fed back from a state estimator which will be explained later, and
\begin{eqnarray}
\mathbf{L}=\left(\mathbf{B}^T\mathbf{S}\mathbf{B}+\mathbf{P}\right)^{-1}\mathbf{B}^T\mathbf{S}\mathbf{A},
\end{eqnarray}
and the matrix $\mathbf{x}$ satisfies the algebraic Riccati Equation, which is given by
\begin{eqnarray}
\mathbf{S}&=&\mathbf{A}^T\bigg(\mathbf{S}-\mathbf{S}\mathbf{B}\nonumber\\
&\times&\left(\mathbf{B}^T\mathbf{S}\mathbf{B}+\mathbf{P}\right)^{-1}\mathbf{B}^T\mathbf{S}\bigg)\mathbf{A}+\mathbf{Q}.
\end{eqnarray}

\section{Trustiness Evaluation}\label{sec:evaluation}
\begin{figure}
  \centering
  \includegraphics[scale=0.5]{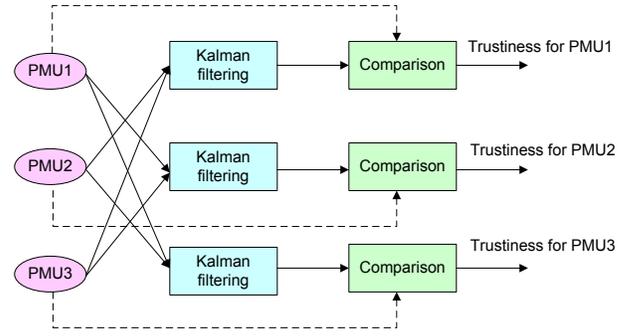}
  \caption{An illustration of the Kalman filtering based trustiness evaluation.}\label{fig:Kalman}
\end{figure}

In this section, we propose a mechanism for evaluating the trustiness of each sensor. The essential reason that the controller can evaluate the trustiness of each sensor is that the controller can predict the future state with some uncertainty. If the report from a sensor is significantly deviated from the prediction, then the controller can consider this sensor as unreliable and ignores its reports. The basic principle is to evaluate the trustiness of each PMU by comparing its report with the prediction obtained from the reports of other $N-1$ PMUs. The procedure is illustrated in Fig. \ref{fig:Kalman}.

\subsection{System State Prediction}
In each time slot, the controller can predict the system state in the next time slot, i.e., $\mathbf{x}(t+1)$, according to its own action $\mathbf{u}(t)$ and the current system state $\mathbf{x}(t)$. Since it is possible that there is one attacker, the controller computes $N$ predictions by excluding one sensor in each prediction.

The Kalman filtering can be applied for the system state estimation. According to \cite{Poor1994}, the system state $\mathbf{x}(t+1)$ is Gaussian distributed. When the observation $y_n$ from PMU $n$ is excluded, the expectation of the system state is given by
\begin{eqnarray}
\mathbf{x}^n(t+1|t)=\mathbf{A}_t\mathbf{x}^n(t|t),
\end{eqnarray}
where
\begin{eqnarray}
\mathbf{x}^n(t|t)=\mathbf{x}^n(t|t-1)+\mathbf{K}^n(t)\left(\mathbf{y}-\mathbf{C}^n\mathbf{x}(t|t-1)\right),
\end{eqnarray}
where $\mathbf{C}^n$ is obtained by removing the $n$-th row from the observation matrix $\mathbf{C}$, and
\begin{eqnarray}
\mathbf{K}^n(t)=\mathbf{\Sigma}(t|t-1)\left(\mathbf{C}^n\mathbf{\Sigma}^n(t|t-1)\left(\mathbf{C}^n\right)^T+\sigma_n^2\mathbf{I}\right)^{-1},
\end{eqnarray}
and covariance matrix given by
\begin{eqnarray}
\mathbf{\Sigma}^n(t|t)=\mathbf{\Sigma}^n(t|t-1)-\mathbf{K}^n_t\mathbf{C}^n(t)\mathbf{\Sigma}(t|t-1),
\end{eqnarray}
where
\begin{eqnarray}
\mathbf{\Sigma}^n(t+1|t)=\mathbf{A}\mathbf{\Sigma}^n(t|t)\mathbf{A}^T+\mathbf{B}\mathbf{Q}\mathbf{B}^T.
\end{eqnarray}

\subsection{Trustiness Computation}
Using the prediction obtained from the reports of PMUs except PMU $n$, the {\em a posteriori} probability of $y_n(t)$, i.e., $p(y_n(t)|\mathbf{y}_{-n}(0:t))$ (here $\mathbf{y}_{-n}$ means the observations excluding that of PMU $n$), is Gaussian distributed with the expectation given by
\begin{eqnarray}\label{eq:mean}
u^n(t)=\mathbf{c}_n\mathbf{x}^n(t|t),
\end{eqnarray}
where $\mathbf{c}_n$ is the $n$-th row in matrix $\mathbf{C}$, and variance
\begin{eqnarray}\label{eq:variance}
\sigma^n(t)=\mathbf{c}_n\mathbf{\Sigma}^n(t|t)\mathbf{c}_n^T.
\end{eqnarray}

We denote by $T_n$ the type of PMU $n$. $T_n=A$ if PMU $n$ is an attacker; otherwise $T_n=H$. Then, we define the suspicious level of PMU $n$ as the following conditional probability, which is given by
\begin{eqnarray}
\pi_n(t)\triangleq P(T_n=A|\mathbf{y}(0:t)).
\end{eqnarray}
The trustiness can be defined as $1-\pi_n(t)$.

The challenges for computing the suspicious level is the unknown attacking strategy\footnote{It is possible that an attacker is captured and the strategy is known to the defender side. However, such an assumption is too strong for most systems.}. We first assume that there must be an attacker. Using the Bayesian rule, we have
\begin{eqnarray}\label{eq:trust}
\pi_n(t)&=&\frac{P(\mathbf{y}(0:t)|T_n=A)}{\sum_{m=1}^NP(\mathbf{y}(0:t)|T_m=A)}\nonumber\\
        &=&\frac{1}{1+\sum_{m\neq n}\frac{P(\mathbf{y}(0:t)|T_m=A)}{P(\mathbf{y}(0:t)|T_n=A)}}\nonumber\\
        &\approx&\frac{1}{1+\sum_{m\neq n}\frac{\prod_{s=0}^tP(\mathbf{y}(s)|T_m=A)}{\prod_{s=0}^tP(\mathbf{y}(s)|T_n=A)}},
\end{eqnarray}
where the last approximation is obtained by decomposing the joint distribution into the product of the probabilities in each time slot. Note that this approximation is not rigorous. However, it simplifies the computation and the validity will be demonstrated in the numerical simulations. We do the following further simplification:
\begin{eqnarray}\label{eq:approx}
P(\mathbf{y}(s)|T_n=A)&\approx& P(y_n(s)|T_n=A)\nonumber\\
&&\prod_{k\neq n}P(y_k(s)|T_k=H),
\end{eqnarray}
by assuming the independence among the PMUs. Although this assumption does not hold, it simplifies the analysis. We further assume that $P(y_n(s)|T_n=A)$ is a constant since we have no knowledge about the attacker's strategy. Substituting the approximation in (\ref{eq:approx}) into (\ref{eq:trust}), we obtain
\begin{eqnarray}
\pi_n(t)\approx \frac{\prod_{s=0}^t\frac{1}{P(y_n(s)|T_n=H)}}{\sum_{m=1}^N\prod_{s=0}^t\frac{1}{P(y_m(s)|T_m=H)}}.
\end{eqnarray}
We then approximate the probability $P(y_n(s)|T_n=H)$ by $\mathcal{N}(u^n(t),\sigma^n(t))$.

When it is possible that there is no attacker, it is easy to repeat the above procedure and obtain the following approximation for the suspicious level of PMU $n$, which is given by
\begin{eqnarray}
\pi_n(t)\approx \frac{\prod_{s=0}^t\frac{1}{P(y_n(s)|T_n=H)}}{L+\sum_{m=1}^N\prod_{s=0}^t\frac{1}{P(y_m(s)|T_m=H)}},
\end{eqnarray}
where $L$ represents the {\em a priori} likelihood that there is no attacker, which is given by
\begin{eqnarray}
L=\frac{P(\mbox{there is no attacker})}{P(\mbox{there is one attacker})}.
\end{eqnarray}
Obviously, the large $L$ is, the less sensitive the controller is to possible attackers.

\subsection{Secure Control Based on Trustiness}
One approach to handle the attacker is to omit its reports once determining that it is an attacker. Alternatively, we propose a heuristic approach for the control based on the trustiness values, called {\em weighted prediction}. Suppose that we still use the LQR control. Then, the control action taken at time $t$ is given by
\begin{eqnarray}
\mathbf{u}(t)=-\mathbf{L}\hat{\mathbf{x}}(t,\left\{P(T_n=A|\mathbf{y})\right\}_n),
\end{eqnarray}
where the system state estimation is a function dependent on the suspicious levels of different PMUs. We set
\begin{eqnarray}
\hat{\mathbf{x}}(t,\left\{P(T_n=A|\mathbf{y})\right\}_n)=\frac{\sum_{n=1}^NP(T_n=A)|\mathbf{y})\hat{\mathbf{x}}^n(t)}{\sum_{n=1}^NP(T_n=A)|\mathbf{y})},
\end{eqnarray}
where the estimation is the weighted sum of the system state estimations of different excluded PMUs. When the suspicious level of PMU $n$ is high, the corresponding system state estimation $\hat{\mathbf{x}}^n(t)$, which excludes the reports from PMU $n$, will dominate (recall that $\mathbf{x}^n(t)$ is obtained by excluding the reports from PMU $n$).

\subsection{Algorithm Summary}
The proposed algorithms are summarized in Procedure \ref{alg:trustiness}.

\begin{procedure}[h]
\begin{center}
\caption{Procedure of The Trustiness Computation and Control}\label{alg:trustiness}
\begin{algorithmic}[1]
    \small
    \FOR{Each time slot $t$}
        \FOR{PMU $n$, $n=1,2,...,N$}
            \STATE{Exclude the report from PMU $n$, i.e. $y_n(t)$.}
            \STATE{Carrying out the Kalman filter for the observation with $y_n(t)$ excluded.}
            \STATE{Compute the expectation and variance using (\ref{eq:mean}) and (\ref{eq:variance}).}
            \STATE{Compute the corresponding probability $P(y_m(s)|T_m=H)$.}
        \ENDFOR
        \STATE{Compute the suspicious levels.}
        \STATE{Apply the weighted system state estimation for the LQR control.}
    \ENDFOR
\end{algorithmic}
\end{center}
\end{procedure}

\section{Numerical Simulations}\label{sec:numerical}
In this section, we use numerical simulations to demonstrate the proposed trustiness system in smart grid.

\subsection{Linear Model}
We adopt the linear model analyzed in Example 6.2 of \cite{Anderson2003}, in which the system is described using the following continuous-time linear dynamics:
\begin{eqnarray}
\dot{\mathbf{x}}(t)=-\mathbf{M}^{-1}\mathbf{K}\mathbf{x}-\mathbf{M}^{-1}\mathbf{u}+\mathbf{w},
\end{eqnarray}
where the matrix $\mathbf{M}$ is given by
\begin{small}
\begin{eqnarray}
&&\mathbf{M}=\nonumber\\
&&\left(
             \begin{array}{ccccccc}
               2.1 & 1.55 & 1.55 & 0 & 0 & 0 & 0 \\
               1.55 & 1.651 & 1.55 & 0 & 0 & 0 & 0 \\
               1.55 & 1.55 & 1.605 & 0 & 0 & 0 & 0 \\
               0 & 0 & 0 & 2.04 & 1.49 & 0 & 0 \\
               0 & 0 & 0 & 1.49 & 1.526 & 0 & 0 \\
               0 & 0 & 0 & 0 & 0 & -1786.9 & 0 \\
               0 & 0 & 0 & 0 & 0 & 0 & 1 \\
             \end{array}
           \right),\nonumber
\end{eqnarray}
\end{small}
and the matrix $\mathbf{K}$ is given in (\ref{eq:mat_K}) (at the top of the next page). The details of the model can be found in \cite{Anderson2003}.
\setcounter{equation}{21}

Since we discuss the discrete-time model in this paper, we approximate the continuous-time model by setting a small time step $\Delta t$, which is given by
\begin{eqnarray}
&&\frac{\mathbf{x}((n+1)\Delta t)-\mathbf{x}(n\Delta t)}{\Delta t}\nonumber\\
&\approx& -\mathbf{M}^{-1}\mathbf{K}\mathbf{x}(n\Delta t)-\mathbf{M}^{-1}\mathbf{u}(n\Delta t)+\mathbf{w}(n\Delta t).
\end{eqnarray}
Therefore, we assume the following discrete-time model:
\begin{eqnarray}
\mathbf{x}(n+1)&=&\left(\mathbf{I}-\Delta t\mathbf{M}^{-1}\mathbf{K}\right)\mathbf{x}(n)\nonumber\\
&-&\Delta t\mathbf{M}^{-1}\mathbf{u}(n)+\Delta t\mathbf{w}(n),
\end{eqnarray}
where we ignore the step $\Delta t$ in the index.

We assume that the PMUs can observe the system state directly, i.e., $\mathbf{C}=\mathbf{I}$, each PMU for one dimension. We further assume that PMU 1 is malicious while all other PMUs are honest.

\newcounter{mytmpeqncnt}
\begin{figure*}[!t]
\normalsize \setcounter{mytmpeqncnt}{\value{equation}}
\setcounter{equation}{20}
\begin{eqnarray}\label{eq:mat_K}
\mathbf{K}=\left(
             \begin{array}{ccccccc}
               0.0211 & 0 & 0 & 2.04 & 1.49 & 1.43 & -1.025 \\
               0 & 0.0007 & 0 & 0 & 0 & 0 & 0 \\
               0 & 0 & 0.0131 & 0 & 0 & 0 & 0 \\
               -2.1 & -1.55 & -1.55 & 0.0211 & 0 & -1.039 & -1.397 \\
               0 & 0 & 0 & 0 & 0.054 & 0 & 0 \\
               -0.014 & -0.362 & -0.362 & -1.428 & -0.79 & 0 & 0 \\
               0 & 0 & 0 & 0 & 0 & -1 & 0 \\
             \end{array}
           \right).
\end{eqnarray}
\hrulefill \vspace*{4pt}
\end{figure*}
\setcounter{equation}{23}

\subsection{Evolution of Suspicious Level}
\begin{figure}
  \centering
  \includegraphics[scale=0.4]{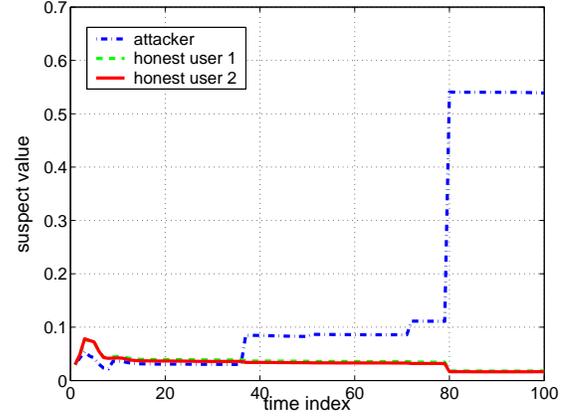}
  \caption{The trace of suspicious level of the attacker and two honest PMUs: random report case.}\label{fig:attack}
\end{figure}

\begin{figure}
  \centering
  \includegraphics[scale=0.4]{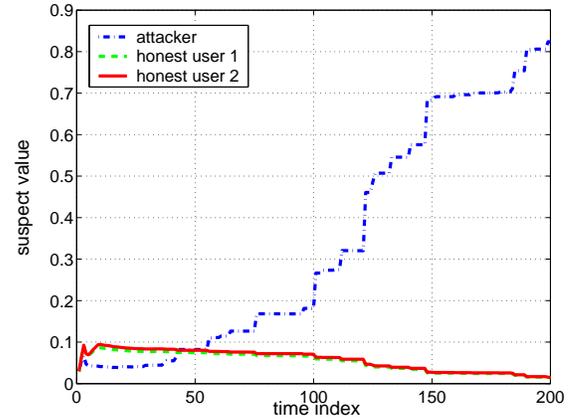}
  \caption{The trace of suspicious level of the attacker and two honest PMUs: deliberate noise case.}\label{fig:attack2}
\end{figure}

In Figures \ref{fig:attack} and \ref{fig:attack2}, the evolution of the suspicious level is shown for the attacker and two honest PMUs. In Fig. \ref{fig:attack}, we assume that the report of the attacker is a Gaussian random variable with zero mean and variance 0.1. The attacker decides to attack or not to attack with probability 0.2 (called {\em attack frequency}). We observe that there is some fluctuation at the beginning. Then, at the times when the attacker launches attacks, the suspicious level of the attacker increases significantly. After around 80 time slots, the attacker can be well distinguished from the two honest PMUs. In Fig. \ref{fig:attack2}, we assume that the attacker attaches a Gaussian noise with zero expectation and variance 0.1 to the observation. The attack frequency is increased to 0.5. We observe that the suspicious level increases more smoothly.

\subsection{Detection Delay and False Alarm}
\begin{figure}
  \centering
  \includegraphics[scale=0.4]{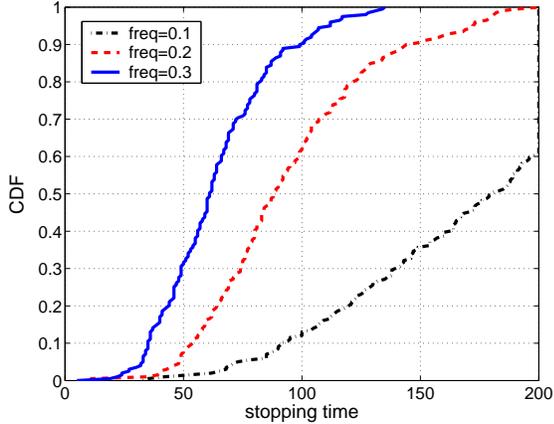}
  \caption{CDF curves of the time of claiming the detection.}\label{fig:CDF}
\end{figure}

In Fig. \ref{fig:CDF}, we plot the cumulative distribution function (CDF) curves of the time when the controller claims that an attacker is detected. We assume that the controller claims the attacker when the suspicious level is larger than 0.7. The cases of false alarms are excluded (note that a false alarm is defined as the event that a honest PMU is claimed to be an attacker). We change the attack frequencies from 0.1 to 0.3. The linear system runs for 200 time slots. We observe that, when the attack frequency is 0.1, the attacker is not detected within the 200 time slots in around 40\% cases. When the attack frequency is increased, the attacker will be detected faster.

\begin{figure}
  \centering
  \includegraphics[scale=0.4]{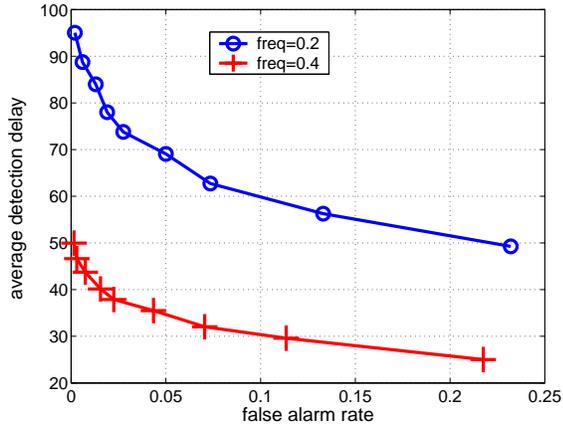}
  \caption{ROC curves (detection delay and false alarm) for different attack frequencies.}\label{fig:ROC}
\end{figure}

In Fig. \ref{fig:ROC}, we plot the receiver operation characteristic (ROC) curves, which show the average delay of detection (excluding the false alarms) and false alarm rate. The attack frequencies are 0.2 and 0.4, respectively. Again, we use the threshold of 0.7 for the suspicious level. Obviously, the detection delay increases when the attack frequency becomes larger with a fixed false alarm rate.

\subsection{Comparison of Cost}
\begin{figure}
  \centering
  \includegraphics[scale=0.4]{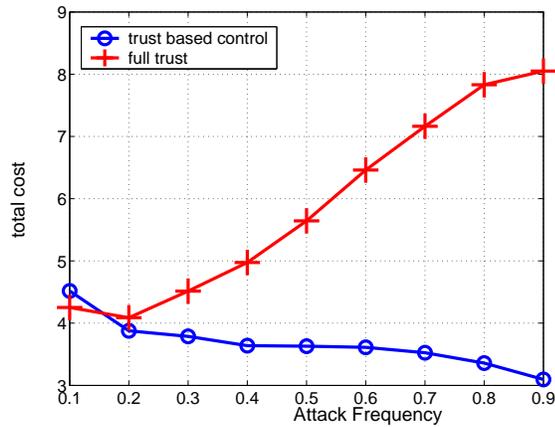}
  \caption{The average cost versus different attack probabilities.}\label{fig:cost1}
\end{figure}

\begin{figure}
  \centering
  \includegraphics[scale=0.4]{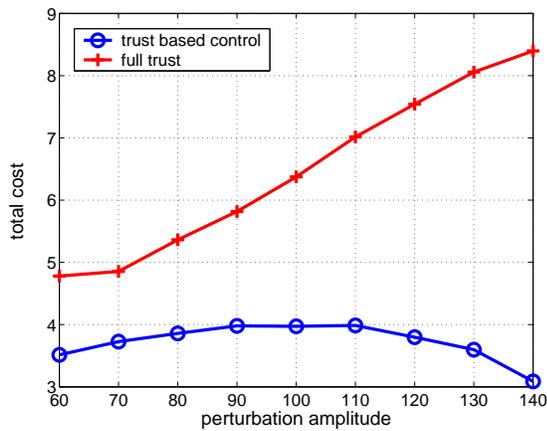}
  \caption{The average cost versus different attack amplitudes.}\label{fig:cost2}
\end{figure}

In Figures \ref{fig:cost1} and \ref{fig:cost2}, the cost averaged over 2000 time slots and 100 realizations is shown. We assume that $\mathbf{Q}=\mathbf{I}$ and $\mathbf{P}=0.01\mathbf{I}$ in (\ref{eq:LQG}), i.e., we pay much more attention to the norm of the system state. We compare the costs using the weighted system state estimation in Section \ref{sec:evaluation} and the system without any counter measure for the attacker (i.e., full trust to each PMU).

In our test, we found that a small amplitude attack (e.g., the attacker uses the same attack as in Fig. \ref{fig:CDF}) causes very small impact on the system. This is because that the Kalman filter has certain inherent robustness since the attack can be partially mitigated by observations from other PMUs. Therefore, we assume that the report of PMU1 is the sum of the original report and a strong noise with a large variance, which is called {\em attack amplitude}. In Fig. \ref{fig:cost1}, we assume that the attack amplitude is 100 and change the attack probabilities. We observe that, as the attack probabilities increases, the average cost of the full trust case is significantly increased. Meanwhile, the total cost is decreased as the attack frequency increases. A possible reason is that a higher attack frequency may cause a more rapid degradation of the trustiness of the attacker. The total cost is also shown for different attack amplitudes in Fig. \ref{fig:cost2}. Again, the total cost increases as the attack amplitude is increased in the full trust case. In a contrast to Fig. \ref{fig:cost1}, the average cost of the weighted system state case is not a monotonic function of the attack amplitude. The reason could be: when the attack amplitude is small, the attacker causes little damage to the system; when the amplitude is large, the controller can detect the attacker early and avoid the cost in later time slots.

\section{Conclusions}\label{sec:conclusion}
In this paper, we have analyzed the possible spoof attack on the smart grid system, in which the attacker can intercept and true report and send its faked report to the controller, which can cause severe damage to the power grid. We have proposed a trustiness system for the controller, in which multiple Kalman filtering processes, with each PMU excluded, are used to cross check the suspiciousness of each PMU. The suspicious levels are then used as weights for the system state estimation for the LQR control. Numerical simulations have shown that the attacker can be effectively detected and the weighted system prediction approach significantly outperforms the system unaware of possible attacks.

Our future work includes the following aspects:
\begin{itemize}
\item The multiple attacker case.
\item The optimal attack strategy of the attacker against the proposed trustiness system.
\end{itemize}


\begin{thebibliography}{11}
\bibitem{Anderson2003}
P. M. Anderson and A. A. Fouad, {\em Power System Control and Stability}, 2nd edition, IEEE Press and Wiley-Interscience, 2003.


\bibitem{Amin2009}
S. Amin, A. A. C\'{a}rdenas and S. Sastry, ``Safe and secure networked control systems under denial-of-service attacks,'' {\em Lecture Notes in Computer Science}, Springer, 2009.


\bibitem{ISO2009}
ISO New England Inc., {\em Overview of the Smart Grid: Policies, Initiatives and Needs}, Feb. 17, 2009.

\bibitem{Cardenas2008_1}
A. A. C\'{a}rdenas, S. Amin and S. Sastry, ``Secure control: Towards survivable cyber-physical systems,'' in {\em Proc. of the 28th International Conference on Distributed Computing Systems Workshops}, 2008.

\bibitem{Cardenas2008_2}
A. A. C\'{a}rdenas, S. Amin and S. Sastry, ``Research challenges for the security of control systems,'' in {\em Proc. of the 3rd Conference on Hot Topics in Security}, 2008.

\bibitem{Josang2007}
A. J{\o}sang, R. Ismail and C. Boyd, ``A survey of trust and reputation systems for online service provision,'' {\em Decision Support Systems}, vol. 43, pp.618--644, March 2007.

\bibitem{Kwakernaak1972}
H. Kwakernaak and R. Sivan, {\em Linear Optimal Control Systems}, 1st Edition, Wiley-Interscience, 1972.

\bibitem{Matveev2009}
A. S. Matveev and A. V. Savkin, {\em Estimation and Control Over Communication Networks}, Birkh\:{a}user, 2009.

\bibitem{Poor1994}
H. V. Poor, {\em An Introduction to Signal Detection and Estimation}, 2nd edition, Springer, 1994.


\bibitem{SCADA2007}
R. Lemos, ``SCADA system makers pushed toward security,'' SecurityFocus, 2007.

\bibitem{Stumpt2007}
F. Stumpt, M. Benz, M. Hermanowski and C. Eckert, ``An approach to a trustworthy system architecture using virtualization,'' Lecture Notes in Computer Science, 2007.



\bibitem{Taylor2005}
C. W. Taylor, D. C. Erickson, K. E. Martin, R. E. Wilson and V. Venkatasubramanian, ``WACS-Wide-are stability and voltage control system: R\&D and online demonstration,'' {\em Proceedings of the IEEE}, pp.892--906, May 2005.



\bibitem{Wei2010}
D. Wei, Y. Lu, M. Jafari, P. Skare and K. Rohde, ``An integrated security system of protecting smart grid against cyber attacks,'' in {\em Proc. of Innovative Smart Grid Technologies Conference}, 2010.


\end{thebibliography}
\end{document}